\documentstyle[epic,eepic]{article}

\begin{document}
\vspace*{15mm}
\begin{center}
{\large \bf Supersymmetric Nambu-Jona-Lasinio Model\\[2mm]
in an External Gravitational Field}
\end{center}
\vspace{8mm}
\begin{center}
I.~L.~Buchbinder
\footnote{e-mail : josephb@tspi.tomsk.su}, \\
{Dept. Theor. Phys., Tomsk Pedagogical University,
634041 Tomsk, Russia,}\\[4mm]
T.~Inagaki
\footnote{e-mail : inagaki@icrr.u-tokyo.ac.jp},\\
{Institute for Cosmic Ray Research, University of Tokyo,
Tanashi, Tokyo 188, Japan,}\\[4mm]
S.~D.~Odintsov
\footnote{e-mail : sergei@ecm.ub.es, odintsov@quantum.univalle.edu.co},\\
{Dept. Theor. Phys., Tomsk Pedagogical University,
634041 Tomsk, Russia\\
and\\
Department de Fisica, Universidad del Valle,\\
A.A.25360 Cali, Colombia}\\[10mm]
\end{center}
\begin{abstract}
We investigate the effect of an external gravitational fields
to the chiral symmetry breaking in the SUSY (supersymmetric) 
NJL (Nambu-Jona-Lasinio) model non-minimally interacting with 
external supergravity.
Evaluating the effective potential in the leading order of the
$1/N_{c}$-expansion and in the linear curvature approximation
it is found the possibility
of the chiral symmetry breaking in the SUSY NJL model
in an external gravitational fields.
In the broken phase the dynamically generated mass is 
analytically and numerically calculated.
\end{abstract}
\vglue 25mm
PACS : 04.62.+v, 04.65.+e, 11.30.Qc, 12.60.Jv
\newpage
\baselineskip 25pt

It is quite old idea \cite{NJL}
that dynamical symmetry breaking in quantum field theory
maybe realized without elementary scalars.
For example, the chiral symmetry breaking may occur
due to formation of the vacuum condensate of composite 
fermion bound states.
Such bound states may play the role of elementary
scalar fields in the standard model, or the role
of the inflaton in the inflationary models
of the early universe where gravitational
effects are quite essential.

Recently, there was some activity (see Ref. \cite{rev}
for a review) on the study of dynamical symmetry breaking
in four-fermion models in curved spacetime.
The phase structure of such a theory has been carefully
investigated.
Depending on the model and spacetime under consideration
the possibility of curvature induced phase transition has
been found.
The dynamical generation of fermion mass has been discussed
in all detail.
Note, however, that in Ref. \cite{rev}
only non-supersymmetric four-fermion models
under the action of the external gravitational field have
been discussed.

The interesting question is what happens with the SUSY 
NJL model in an external (super) gravitational field.
It is known that in flat spacetime \cite{snjl}
SUSY NJL model does not show the dynamical symmetry
breaking.
Hence, could we expect that external (super) gravitational 
field could be the source of phase transitions in SUSY
NJL model?
In the present paper we will make the attempt to answer above
question.

We will start with the action of SUSY NJL model
in an external supergravitational background.
This model can be considered
as local supersymmetry generalization of the model given
in Ref. \cite{snjl}:
\begin{equation}
\begin{array}{rcl}
     S&=&\displaystyle\int d^{8}z E^{-1}
     \left[\bar{Q}Q+\bar{Q}^{c}Q^{c}
           +G(\bar{Q}^{c}Q)(\bar{Q}Q^{c})
           +\bar{\xi}_{1}(\bar{Q}Q)
           +\bar{\xi}_{2}(\bar{Q}^{c}Q^{c})\right],
\label{act:org}
\end{array}
\end{equation}
where chiral superfields $Q^{\alpha}$, $Q^{c}_{\alpha}$
carry the color index $\alpha =1, \cdots, N_{c}$
and belong to the representations of $SU(N_{c})$,
$E=\mbox{Ber} E^{A}_{M}$, $E^{A}_{M}$ is supertetrade,
$\bar{\xi}_{1}$ and $\bar{\xi}_{2}$ are non-minimal
coupling constants of SUSY NJL model
with external supergravity.
%Non-minimal interaction of quantum field theory with 
%external (super)gravity has been studied in all detail 
%in book \cite{book}.
We follow the notation of book \cite{sgra}.
Note also that there exists more general
form of non-minimal interaction above SUSY theory
with supergravity but we consider only a simplest
variant.

After the standard introduction of auxiliary 
superfields\footnote{
We use the term auxiliary superfields in a sense 
acceptable in Nambu-Jona-Lasinio and Gross-Neveu models. Auxiliary fields 
here mean the fields what are introduced to simplify a form of 
interactioin and define the composite fields the average values of which 
play a role of order parameters. These auxiliary (super) fields have no 
relation to auxiliary component fields which are used in supersymmetric 
field theories.
}
and rewriting the action (\ref{act:org}) in component
fields we will limit ourselves to purely gravitational
background.
In the leading order of the $1/N_{c}$-expansion
the spinor auxiliary fields$^{4}$ are dropped away
since they may contribute only to
next-to-leading order terms in the $1/N_{c}$-expansion.
Then the action to start with takes the form:
\begin{equation}
\begin{array}{rcl}
     S&=&\displaystyle\int d^{4}x\sqrt{-g}\biggl[
         -\phi^{\dag}(\nabla^{\mu}\nabla_{\mu}
         +\rho^{2}+\xi_{1} R)\phi
         -{\phi^{c}}^{\dag}(\nabla^{\mu}\nabla_{\mu}
         +\rho^{2}+\xi_{2} R)\phi^{c}\\
       &&\displaystyle +\bar{\psi}(i\gamma^{\mu}\nabla_{\mu}
         -\rho)\psi-\frac{1}{G}\rho^{2}\biggr].      
\end{array}
\label{s:snjl2}
\end{equation}
where $\rho^{2}=\sigma^{2}+\pi^{2}$ is an auxiliary
scalar as in the original NJL model,
$\psi$ is $N_{c}$ component Dirac spinor,
$\xi_{1}=(1+\bar{\xi}_{1})/6$,
$\xi_{2}=(1+\bar{\xi}_{2})/6$.
The minimal interaction with external super gravity 
corresponds to $\xi_{1}=\xi_{2}=1/6$.
The start point action has the chiral symmetry.
If the auxiliary field $\rho$ develops the non-vanishing 
vacuum expectation value, $\langle\rho\rangle =m \neq 0$,
the fermion $\psi$ and the scalar $\phi$ acquire
the dynamical mass $m$ and the chiral symmetry is
eventually broken.

To find the phase structure of the model given by
action (\ref{s:snjl2}) we introduce an effective 
potential.
To evaluate the effective potential ones start with 
generating functional of Green functions. 
\begin{equation}
\begin{array}{rcl}
     Z&=&\displaystyle\int {\cal D}\psi{\cal D}\bar{\psi}
         {\cal D}\rho\ e^{iS}\\
      &=&\displaystyle\int {\cal D}\rho\frac{\mbox{Det}
         (i\gamma^{\mu}\nabla_{\mu}-\rho)}
         {\mbox{Det}(\nabla^{\mu}\nabla_{\mu}
         +\rho^{2}+\xi_{1} R)
         (\nabla^{\mu}\nabla_{\mu}
         +\rho^{2}+\xi_{2} R)}
         \exp i\int d^{4}x\sqrt{-g}
         \left(-\frac{1}{G}\rho^{2}\right)\\
      &=&\displaystyle
         \int {\cal D}\rho\exp i\biggl[\int d^{4}x\sqrt{-g}
         \left(-\frac{1}{G}\rho^{2}\right)
         -i\mbox{ln Det}(i\gamma^{\mu}\nabla_{\mu}-\rho)\\
       &&\displaystyle +i\mbox{ln Det}(\nabla^{\mu}\nabla_{\mu}
         +\rho^{2}+\xi_{1} R)
         +i\mbox{ln Det}(\nabla^{\mu}\nabla_{\mu}
         +\rho^{2}+\xi_{2} R)
         \biggr].
\end{array}
\end{equation}
An internal line of the $\rho$-propagator has no contribution
in the leading order of the $1/N_{c}$-expansion.
Assuming that the $\rho$ is slowely variating field and 
applying the $1/N_{c}$-expansion method the effective potential 
for $\rho$ is found to be
\begin{equation}
     V(\rho)=\frac{1}{G}\rho^{2}
            +i \mbox{tr}\ln(i\gamma^{\mu}\nabla_{\mu}-\rho)
            -i \mbox{tr}\ln(\nabla^{\mu}\nabla_{\mu}
                            +\rho^{2}+\xi_{1} R)
            -i \mbox{tr}\ln(\nabla^{\mu}\nabla_{\mu}
                            +\rho^{2}+\xi_{2} R)
            +{\cal O}\left(\frac{1}{N}\right).
\end{equation}
Using the two-point Green functions the effective potential is
rewritten as
\begin{equation}
     V(\rho)=\frac{1}{G}\rho^{2}
            -i \mbox{tr} \int^{\rho}_{0}ds\ S(x,x;s)
            -2i \int^{\rho}_{0}sds\ [G_{1}(x,x;s)
                                    +G_{2}(x,x;s)]
            +{\cal O}\left(\frac{1}{N}\right),
\label{epot:snjl}
\end{equation}
where $S(x,x;s)$ and $G_{i}(x,x;s)$ are the spinor and scalar
two-point functions respectively satisfying the equations
\begin{equation}
     (i\gamma^{\mu}\nabla_{\mu}-s)S(x,y;s)
     =\frac{1}{\sqrt{-g}}\delta^{4}(x,y),
\end{equation}
\begin{equation}
     (\nabla^{\mu}\nabla_{\mu}+s^{2}+\xi_{i} R)G_{i}(x,y;s)
     =\frac{1}{\sqrt{-g}}\delta^{4}(x,y).
\end{equation}
Here it should be noted that the effective potential 
(\ref{epot:snjl})
is normalized as $V(0)=0$.

We will evaluate the effective potential taking into
account the terms up to linear curvature and
using local momentum representation of propagators
(see Ref. \cite{rev,book} for a review).
The scalar and spinor two-point functions in this approximation
have the forms
\begin{equation}
\begin{array}{rcl}
     \displaystyle S(x,x;s)
     &=&\displaystyle \int
        \frac{d^{4}p}{(2\pi)^{4}}\left[
        \frac{\gamma^{a}p_{a}+s}{p^{2}-s^{2}}
        -\frac{1}{12}R\frac{\gamma^{a}p_{a}+s}{(p^{2}-s^{2})^{2}}
     \right.\\
     &&\displaystyle\left.+\frac{2}{3}{R}^{\mu\nu}p_{\mu}p_{\nu}
     \frac{\gamma^{a}p_{a}+s}{(p^{2}-s^{2})^{3}}
     +\frac{1}{4}\gamma^{a}\sigma^{cd}{R}_{cda\mu}p^{\mu}
     \frac{1}{(p^{2}-s^{2})^{2}}
     \right]+{\cal O}(R_{;\mu},R^{2})\, .
\end{array}
\label{2pf:spinor}
\end{equation}
\begin{equation}
     G_{i}(x,x;s)=\int\frac{d^{4}p}{(2\pi)^{4}}\left[
     -\frac{1}{p^{2}-s^{2}}
     +\left(\frac{1}{3}-\xi_{i} \right)
     \frac{R}{(p^{2}-s^{2})^{2}}
     -\frac{2}{3}\frac{{R}^{\mu\nu}p_{\mu}p_{\nu}}{(p^{2}-s^{2})^{3}}
     \right]
     +{\cal O}(R_{;\mu},R^{2})\, .
\label{2pf:scalar}
\end{equation}

Inserting Eqs.(\ref{2pf:spinor}) and (\ref{2pf:scalar})
into Eq.(\ref{epot:snjl}) the effective potential reads
\begin{equation}
\begin{array}{rcl}
     \displaystyle V(\rho)&=&
     \displaystyle \frac{1}{G}\rho^{2}
      -4 i \int^{\rho}_{0}sds
       \int \frac{d^{4}p}{(2\pi)^{4}}\left[
       \frac{1}{p^{2}-s^{2}}
       -\frac{1}{12}R\frac{1}{(p^{2}-s^{2})^{2}}
       +\frac{2}{3}R^{\mu\nu}p_{\mu}p_{\nu}
       \frac{1}{(p^{2}-s^{2})^{3}}\right]\\
     &&\displaystyle +2 i \int^{\rho}_{0}sds
       \int \frac{d^{4}p}{(2\pi)^{4}}\left[
       \frac{2}{p^{2}-s^{2}}
       -\left(\frac{2}{3}-\xi_{1}-\xi_{2} \right)
       R\frac{1}{(p^{2}-s^{2})^{2}}
       +\frac{4}{3}R^{\mu\nu}p_{\mu}p_{\nu}
       \frac{1}{(p^{2}-s^{2})^{3}}\right]\\
     &=&\displaystyle \frac{1}{G}\rho^{2}
       +2 i  \left(\frac{1}{6}-\frac{2}{3}+\xi_{1}+\xi_{2}\right)R
       \int^{\rho}_{0}sds
       \int \frac{d^{4}p}{(2\pi)^{4}}
       \frac{1}{(p^{2}-s^{2})^{2}}
       +{\cal O}\left(\frac{1}{N}\right).
\end{array}
\label{epot:snjl:wc}
\end{equation}
Thus for $\xi_{1}+\xi_{2}=1/2$ the fermion loop contribution
is cancelled with the boson loop contribution.
In this case the chiral symmetry is not broken
down even in curved spacetime.
To evaluate the integration in Eq.(\ref{epot:snjl:wc}) 
we perform the Wick rotation $p^{0}\rightarrow ip^{0}$
\begin{equation}
     I= i  R \int^{\rho}_{0}sds
       \int \frac{d^{4}p}{(2\pi)^{4}}
       \frac{1}{(p^{2}-s^{2})^{2}}
     \rightarrow - R \int^{\rho}_{0}sds
       \int \frac{d^{4}p}{(2\pi)^{4}}
       \frac{1}{(p^{2}+s^{2})^{2}}.
\label{epot:snjl:lhs}
\end{equation}
Applying the Schwinger proper time method \cite{Schw} $I$ 
is rewritten as
\begin{equation}
     I= - R \int^{\rho}_{0}sds
      \int \frac{d^{4}p}{(2\pi)^{4}}
      \int^{\infty}_{0}tdt\ e^{-t(p^{2}+s^{2})}
     =\frac{R}{2(4\pi)^{2}}\int^{\infty}_{0}dt\frac{1}{t^{2}}
      \left(e^{-t\rho^{2}}-1\right).
\end{equation}
Since the integration over $t$ is divergent around $t\sim 0$,
we introduce the proper time cut-off $\Lambda$ and find
\begin{equation}
     I\rightarrow\frac{R}{2(4\pi)^{2}}\int^{\infty}_{1/\Lambda^{2}}
       dt\frac{1}{t^{2}}
       \left(e^{-t\rho^{2}}-1\right)
      = \frac{R}{2(4\pi)^{2}}\left[
          \rho^{2}\mbox{Ei}
          \left(-\frac{\rho^{2}}{\Lambda^{2}}\right)
          +\Lambda^{2}
          (e^{-\rho^{2}/\Lambda^{2}}-1)          
          \right],
\end{equation}
where $\mbox{Ei}(-x)$ is the exponential-integral function
which is defined by
\begin{equation}
     \mbox{Ei}(-x)=-\int^{\infty}_{x}dt\frac{e^{-t}}{t}
     =\ln x+\gamma+\sum^{\infty}_{n=1}
        \frac{(-x)^{n}}{n\cdot n!} < 0\ ;\ (x > 0).
\label{pro:ei}
\end{equation}
Hence the effective potential in the leading order of
the $1/N_{c}$-expansion for the supersymmetric NJL model
in curved spacetime reads
\begin{equation}
     V(\rho)= \frac{1}{G}\rho^{2}
     -\frac{R}{(4\pi)^{2}}
          f(\xi_{1},\xi_{2})
          \left[
          \rho^{2}\mbox{Ei}
          \left(-\frac{\rho^{2}}{\Lambda^{2}}\right)
          +\Lambda^{2}
          (e^{-\rho^{2}/\Lambda^{2}}-1)          
          \right],
\label{epot:snjl:wcfin}
\end{equation}
where $f(\xi_{1},\xi_{2})$ is
\begin{equation}
     f(\xi_{1},\xi_{2})=\frac{1}{2}-\xi_{1}-\xi_{2}.
\label{def:f}
\end{equation}

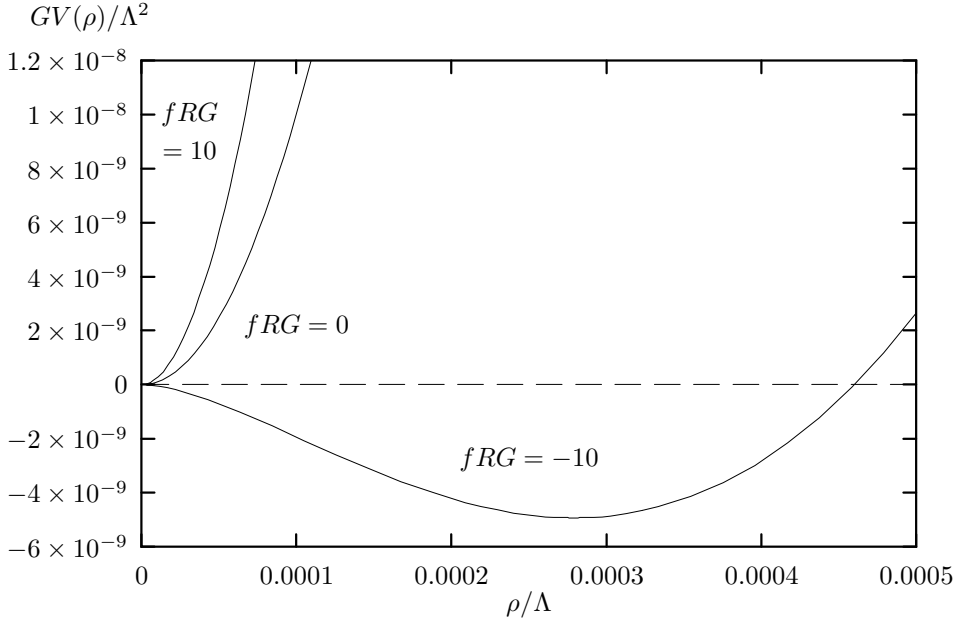
\begin{figure}
% GNUPLOT: LaTeX picture using EEPIC macros
\setlength{\unitlength}{0.240900pt}
\begin{picture}(1500,900)(0,0)
\rm
\put(828,250){\makebox(0,0){$fRG=-10$}}
\put(463,460){\makebox(0,0){$fRG=0$}}
\put(295,792){\makebox(0,0){$fRG$}}
\put(295,735){\makebox(0,0){$=10$}}
\thicklines \path(220,113)(240,113)
\thicklines \path(1436,113)(1416,113)
\put(198,113){\makebox(0,0)[r]{$-6\times 10^{-9}$}}
\thicklines \path(220,198)(240,198)
\thicklines \path(1436,198)(1416,198)
\put(198,198){\makebox(0,0)[r]{$-4\times 10^{-9}$}}
\thicklines \path(220,283)(240,283)
\thicklines \path(1436,283)(1416,283)
\put(198,283){\makebox(0,0)[r]{$-2\times 10^{-9}$}}
\thicklines \path(220,368)(240,368)
\thicklines \path(1436,368)(1416,368)
\put(198,368){\makebox(0,0)[r]{0}}
\thicklines \path(220,453)(240,453)
\thicklines \path(1436,453)(1416,453)
\put(198,453){\makebox(0,0)[r]{$2\times 10^{-9}$}}
\thicklines \path(220,537)(240,537)
\thicklines \path(1436,537)(1416,537)
\put(198,537){\makebox(0,0)[r]{$4\times 10^{-9}$}}
\thicklines \path(220,622)(240,622)
\thicklines \path(1436,622)(1416,622)
\put(198,622){\makebox(0,0)[r]{$6\times 10^{-9}$}}
\thicklines \path(220,707)(240,707)
\thicklines \path(1436,707)(1416,707)
\put(198,707){\makebox(0,0)[r]{$8\times 10^{-9}$}}
\thicklines \path(220,792)(240,792)
\thicklines \path(1436,792)(1416,792)
\put(198,792){\makebox(0,0)[r]{$1\times 10^{-8}$}}
\thicklines \path(220,877)(240,877)
\thicklines \path(1436,877)(1416,877)
\put(198,877){\makebox(0,0)[r]{$1.2\times 10^{-8}$}}
\thicklines \path(220,113)(220,133)
\thicklines \path(220,877)(220,857)
\put(220,68){\makebox(0,0){0}}
%\thicklines \path(342,113)(342,133)
%\thicklines \path(342,877)(342,857)
%\put(342,68){\makebox(0,0){$5\times 10^{-5}$}}
\thicklines \path(463,113)(463,133)
\thicklines \path(463,877)(463,857)
\put(463,68){\makebox(0,0){0.0001}}
%\thicklines \path(585,113)(585,133)
%\thicklines \path(585,877)(585,857)
%\put(585,68){\makebox(0,0){0.00015}}
\thicklines \path(706,113)(706,133)
\thicklines \path(706,877)(706,857)
\put(706,68){\makebox(0,0){0.0002}}
%\thicklines \path(828,113)(828,133)
%\thicklines \path(828,877)(828,857)
%\put(828,68){\makebox(0,0){0.00025}}
\thicklines \path(950,113)(950,133)
\thicklines \path(950,877)(950,857)
\put(950,68){\makebox(0,0){0.0003}}
%\thicklines \path(1071,113)(1071,133)
%\thicklines \path(1071,877)(1071,857)
%\put(1071,68){\makebox(0,0){0.00035}}
\thicklines \path(1193,113)(1193,133)
\thicklines \path(1193,877)(1193,857)
\put(1193,68){\makebox(0,0){0.0004}}
%\thicklines \path(1314,113)(1314,133)
%\thicklines \path(1314,877)(1314,857)
%\put(1314,68){\makebox(0,0){0.00045}}
\thicklines \path(1436,113)(1436,133)
\thicklines \path(1436,877)(1436,857)
\put(1436,68){\makebox(0,0){0.0005}}
\thicklines \path(220,113)(1436,113)(1436,877)(220,877)(220,113)
\put(45,945){\makebox(0,0)[l]{\shortstack{$GV(\rho)/\Lambda^2$}}}
\put(828,23){\makebox(0,0){$\rho/\Lambda$}}
\thinlines \path(220,368)(220,368)(222,368)(224,368)(225,368)
(226,368)(227,368)(230,368)(232,369)(237,370)(242,371)(253,375)
(263,381)(273,388)(283,396)(293,406)(303,417)(313,429)(323,443)
(333,459)(343,476)(353,494)(363,514)(373,535)(383,558)(393,582)
(403,608)(413,635)(423,664)(433,694)(443,725)(453,758)(463,792)
(475,835)(486,877)
\thinlines \path(220,368)(220,368)(222,368)(224,368)(225,368)
(226,368)(228,369)(230,370)(234,371)(238,374)(246,380)(254,388)
(262,399)(270,411)(278,426)(286,442)(294,461)(302,481)(310,504)
(318,528)(327,555)(335,583)(343,613)(351,645)(359,679)(367,715)
(375,753)(383,793)(391,835)(398,877)
\thinlines \path(222,368)(222,368)(224,368)(226,368)(229,367)
(232,367)(235,367)(248,365)(260,363)(273,360)(324,344)(374,325)
(425,303)(475,280)(526,257)(576,236)(627,215)(678,198)(728,182)
(753,176)(779,171)(804,166)(829,163)(842,161)(854,160)(867,159)
(873,159)(880,159)(883,159)(886,159)(889,159)(891,158)(892,158)
(894,158)(896,158)(897,158)(899,158)(900,158)(902,158)(903,158)
(905,158)(908,159)(911,159)(915,159)(918,159)(924,159)(930,159)
(943,160)(956,161)
(981,165)(1006,170)
\thinlines\path(1006,170)(1031,176)(1082,192)(1133,214)
(1183,241)(1234,276)(1284,316)(1335,364)(1385,418)(1436,480)
\thinlines\drawline[-30](222,368)(1436,368)
\end{picture}
\caption{Behaviour of the effective potential at $f(\xi_{1},\xi_{2})
RG=10, 0, -10$.}
\end{figure}

The ground state of the system corresponds to
the minimum of the effective potential.
To find the ground state we calculate the effective potential
numerically with varying the $f(\xi_{1},\xi_{2})RG$.
In Fig. 1 we show the typical behaviour of the effective potential 
(\ref{epot:snjl:wcfin}) at $f(\xi_{1},\xi_{2})RG=10, 0, -10$.

In flat spacetime $R=0$
the effective potential (\ref{epot:snjl:wcfin}) 
takes a very simlpe form of 
quadratic function. It is evident that the minimum of the effective 
potential will be only at $\rho = 0$ for arbitrary quntities of the 
nonminimal coupling parameters. This situation is illustrated by 
Fig. 1 at $f(\xi_{1},\xi_{2})RG=0$.

For $f(\xi_{1},\xi_{2})RG > 0$ taking into account the property 
(\ref{pro:ei}) of the exponential-integral function we see that 
the effective potenial (\ref{epot:snjl:wcfin}) is 
non-negative and takes minimum at $\rho = 0$. This situation is
illustrated by Fig. 1 at $f(\xi_{1},\xi_{2})RG=10$.
%As is seen in Fig.1 the minimum of the 
%effective potential is $\rho =0$ for $R=0$.
%Thus the ground state preserves the chiral symmetry in
%flat spacetime.
%There is only the symmetric phase
%for an arbitrary positive $f(\xi_{1},\xi_{2})RG$.

As is seen in Fig. 1 the chiral symmetry is broken down for
$f(\xi_{1},\xi_{2})RG=-10$.
To understand the phase structure of the SUSY NJL model
for a negative $f(\xi_{1},\xi_{2})RG$
we will analyse the effective potential more precisely.
%Since the symmetric phase is realised
%for an arbitrary $R$ in the case $f(\xi_{1},\xi_{2})=0$,
%we consider the case $f(\xi_{1},\xi_{2})\neq 0$ here.
The dynamically generated mass $m$ of fermion $\psi$ and 
scalar $\phi$ is given by the value of $\rho$ at the
minimum of the effective potential.
Stationary condition for the effective potential
(\ref{epot:snjl:wcfin}) is given by
\begin{equation}
     \frac{\partial V(\rho)}{\partial \rho}=
     \rho\left[\frac{2}{G}-\frac{2R}{(4\pi)^{2}}
     f(\xi_{1},\xi_{2})
     \mbox{Ei}\left(-\frac{\rho^{2}}{\Lambda^{2}}\right)\right]=0.
\label{gap:snjl}
\end{equation}
Thus the dynamical mass $m=\langle\rho\rangle$ satisfies
\begin{equation}
     \frac{16\pi^{2}}{RG}=
     f(\xi_{1},\xi_{2})
     \mbox{Ei}\left(-\frac{m^{2}}{\Lambda^{2}}\right) < 0.
\label{mass:snjl}
\end{equation}
This equation has a solution only for a negative 
$f(\xi_{1},\xi_{2})RG$.
In Fig.2 the dynamically generated mass is plotted 
as a function of $f(\xi_{1},\xi_{2})RG$.
%As is seen in Fig.2 the second order phase 
%transition takes place and
The dynamical mass $m=\langle\rho\rangle$ which 
corresponds to the order parameter smoothly disappears 
as the $f(\xi_{1},\xi_{2})RG$ increases.

\begin{figure}
% GNUPLOT: LaTeX picture using EEPIC macros
\setlength{\unitlength}{0.240900pt}
\begin{picture}(1500,900)(0,0)
\rm
\thicklines \path(220,113)(240,113)
\thicklines \path(1436,113)(1416,113)
\put(198,113){\makebox(0,0)[r]{0}}
\thicklines \path(220,189)(240,189)
\thicklines \path(1436,189)(1416,189)
\put(198,189){\makebox(0,0)[r]{0.0001}}
\thicklines \path(220,266)(240,266)
\thicklines \path(1436,266)(1416,266)
\put(198,266){\makebox(0,0)[r]{0.0002}}
\thicklines \path(220,342)(240,342)
\thicklines \path(1436,342)(1416,342)
\put(198,342){\makebox(0,0)[r]{0.0003}}
\thicklines \path(220,419)(240,419)
\thicklines \path(1436,419)(1416,419)
\put(198,419){\makebox(0,0)[r]{0.0004}}
\thicklines \path(220,495)(240,495)
\thicklines \path(1436,495)(1416,495)
\put(198,495){\makebox(0,0)[r]{0.0005}}
\thicklines \path(220,571)(240,571)
\thicklines \path(1436,571)(1416,571)
\put(198,571){\makebox(0,0)[r]{0.0006}}
\thicklines \path(220,648)(240,648)
\thicklines \path(1436,648)(1416,648)
\put(198,648){\makebox(0,0)[r]{0.0007}}
\thicklines \path(220,724)(240,724)
\thicklines \path(1436,724)(1416,724)
\put(198,724){\makebox(0,0)[r]{0.0008}}
\thicklines \path(220,801)(240,801)
\thicklines \path(1436,801)(1416,801)
\put(198,801){\makebox(0,0)[r]{0.0009}}
\thicklines \path(220,877)(240,877)
\thicklines \path(1436,877)(1416,877)
\put(198,877){\makebox(0,0)[r]{0.001}}
\thicklines \path(423,113)(423,133)
\thicklines \path(423,877)(423,857)
\put(423,68){\makebox(0,0){-10}}
%\thicklines \path(676,113)(676,133)
%\thicklines \path(676,877)(676,857)
%\put(676,68){\makebox(0,0){-15}}
\thicklines \path(929,113)(929,133)
\thicklines \path(929,877)(929,857)
\put(929,68){\makebox(0,0){-5}}
%\thicklines \path(1183,113)(1183,133)
%\thicklines \path(1183,877)(1183,857)
%\put(1183,68){\makebox(0,0){-5}}
\thicklines \path(1436,113)(1436,133)
\thicklines \path(1436,877)(1436,857)
\put(1436,68){\makebox(0,0){0}}
\thicklines \path(220,113)(1436,113)(1436,877)(220,877)(220,113)
\put(45,945){\makebox(0,0)[l]{\shortstack{$m/\Lambda$}}}
\put(828,23){\makebox(0,0){$f(\xi_{1},\xi_{2})R G$}}
\thinlines \path(845,114)(845,114)(806,115)(783,116)(766,117)
(752,118)(731,119)(715,122)(689,126)(670,130)(640,138)(617,146)
(583,161)(556,177)(515,209)(484,241)(458,273)(436,305)(417,336)
(399,368)(383,400)(368,432)(355,464)(342,495)(330,527)(318,559)
(307,591)(297,623)(287,654)(278,686)(268,718)(260,750)(251,782)
(243,813)(235,845)(227,877)
\end{picture}
\caption{Behaviour of the dynamically generated mass $m$
         with varying the $f(\xi_{1},\xi_{2})RG$.}
\end{figure}
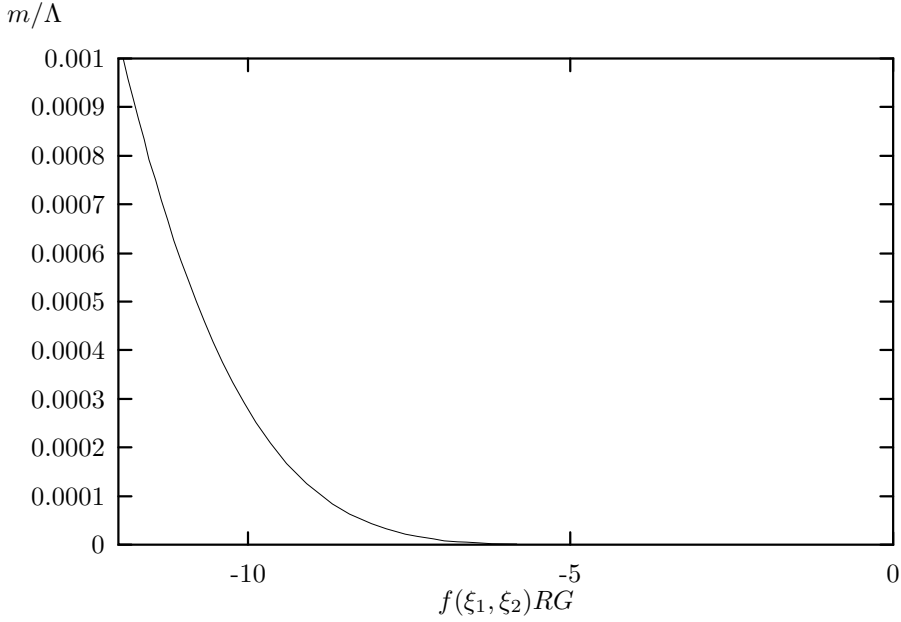

The critical curvature $R_{cr}$ which divides the symmetric
phase and the broken phase is given by the curvature
where the order parameter $m$ disappears.
If we take the limit $m/\Lambda\rightarrow 0$
in Eq.(\ref{mass:snjl}),
we find
\begin{eqnarray}
     RG&=&\frac{16\pi^{2}}{f(\xi_{1},\xi_{2})}
        \left[\ln\frac{m^{2}}{\Lambda^{2}}+\gamma
        +{\cal O}\left(\frac{m^{2}}{\Lambda^{2}}\right)
     \right]^{-1}\nonumber \\
     &\rightarrow& 0\, \,  (f(\xi_{1},\xi_{2})\neq 0).
\end{eqnarray}
Due to the stability of the ground state the coupling $G$
must be real and positive (see Fig.1).
Hence the critical curvature is given by $R_{cr}=0$ and
the chiral symmetry is broken down for an
arbitrary negative $f(\xi_{1},\xi_{2})R$.

For $m/\Lambda\ll 1$ the dynamical mass $m$ is obtained by
\begin{equation}
     m^{2}=\Lambda^{2}\left[
           \exp\left(\frac{16\pi^{2}}{f(\xi_{1},\xi_{2})RG}
           -\gamma\right)
           +{\cal O}\left(\frac{m^{2}}{\Lambda^{2}}\right)
           \right].
\end{equation}
The dynamical mass is suppressed exponentially
for a small negative $f(\xi_{1},\xi_{2})RG$.

Thus it is found that the chiral symmetry is
broken down for an arbitrary negative $f(\xi_{1},\xi_{2})R$
in SUSY NJL model within our approximation.
For $\xi_{1}+\xi_{2} < 1/2$
the broken phase is realised in a spacetime with
an arbitrary negative curvature and the symmetric
phase is realized for an arbitrary positive curvature.
On the other hand there is only the broken phase 
in positive curvature spacetime
and the symmetric phase at a negative
curvature for $\xi_{1}+\xi_{2} > 1/2$.
In both cases the critical curvature is given by 
$R_{cr}=0$.

Note finally that there are few possibilities
to generalize the results of the present work.
First of all, it could be interesting to study
phase structure of SUSY NJL model on non-trivial 
supergravitational background (with non-zero
gravitino).
Second, it could be interesting to investivate
the gauged SUSY NJL model where compositeness conditions
a la Bardeen-Hill-Lindner \cite{bhl} maybe implemented.
That could lead to the formulation of compositeness 
condition for $\xi_{1}$ and $\xi_{2}$
which presumbly should be given by asymptotically
(super) conformal invariant values. \cite{book,bo}

The work by ILB and SDO was supported in part by RFBR 
under the project No 96-02-16017. The work by ILB was 
supported in part by RFBR-DFG under the project 
No 96-02-00180G.

\end{document}